# On the Risk Management with Application of Econophysics Analysis in Central Banks and Financial Institutions.

Dimitri O. Ledenyov and Viktor O. Ledenyov

*Abstract* – The purpose of this research article is to discover how the *econophysics* analysis can complement the *econometrics* models in application to the risk management in the central banks and financial institutions, operating within the nonlinear dynamical financial system. We consider the modern risk management models and show the appropriate techniques to calculate the various existing risks in the finances. We make a few comments on the possible limitations in the models of statistical modeling of volatility such as the *Autoregressive Conditional Heteroskedasticity (GARCH)* model, because of the nonlinearities appearance in the nonlinear dynamical financial systems. We propose that the various types of nonlinearities, which can originate in the financial and economical systems, have to be taken to the detailed consideration during the *Cost of Capital* calculation in the finances and economics. We propose the new **theory of nonlinear dynamic volatilities** and the new **nonlinear dynamic chaos (NDC) volatility model** for the statistical modeling of financial volatility with the aim to determine the *Value at Risk*.

The well designed, perfectly optimized and efficiently operated *financial system* is a main foundation of prosperous developed society with the market economy as it is explained in the political economy in *Landes (1998)*; *Hara (2012)*; *Hirch (1896); Ledenyov V O, Foldvary F (1993-1994)*. The fact is that the present global financial system, which is established due to the evolution of the national and multinational banking in the *UK* and *USA* in *Jones (1993, 2006); Friedman, Schwartz (1971)*; *Rothbard (2002)*, can be characterized by the high volatilities in the capital markets and is classified as the nonlinear dynamical financial system.

The purpose of this article is to discover how the *econophysics* analysis can complement the *econometrics* models in *Engle (1982, 2003); Davidson and MacKinnon (2004)* with the aim to improve the risk management practices in the central banks and other financial institutions. Let us explain that the *econophysics* is an emerging scientific discipline that applies the research concepts and methodologies, that are originated in the field of physics, to understand the nature of problems in the fields of economics and finances, using the nonlinear dynamical analysis, statistical analysis, dynamic chaos analysis and probabilities theory. *Aoyama, Fujiwara, Iyetomi, Sato (2012)* write: "Econophysics was developed by researchers in several fields. In the *1960s*, *Mandelbrot* began to investigate price movements of financial markets and established the important concept of fractals through inspiration provided by the observation of their fluctuations. In the *1980s*, physicists began to investigate statistical properties of human behavior on the basis of socio-economic data. However, the availability of data was limited, and their resolution and coverage were bounded. In spite of these difficulties, the topics investigated by physicists eventually expanded to cover various aspects of financial markets and business activities. They further attempted to develop models to explain collective behavior observed in socio-econo-techno systems in terms of physical concepts, such as scaling, clustering, correlations, and more complicated concepts." We will use the knowledge base, created during our advanced innovative research on the nonlinearities in microwave superconductivity in the microwave electronics and condensed matter physics with the purpose to understand the complex problems in the risk management in the finances in *Ledenyov D O, Ledenyov V O (2012)*.

The simple concept of optimization of behaviour with the application of risk management has a long history in the economics in *Engle (2003)*: "*Markowitz (1952)* and *Tobin (1958)* associated risk with the variance in the value of a portfolio. From the avoidance of risk they derived optimizing portfolio and banking behavior. *Sharpe (1964)* developed the implications, when all investors follow the same objectives with the same information. This theory is called the ***Capital Asset Pricing Model*** or ***CAPM***, and shows that there is a natural relation between expected returns and variance. These contributions were recognized by *Nobel prizes* in *1981* and

*1990. Black and Scholes (1972) and Merton (1973) developed a model to evaluate the pricing of options."*

Let us review the **risk management** practices, which are used to mitigate the risk, going from the principles of diversification, hedging and risk measurements. The actual **risk management concept** is reflected in the **Economic Capital** and **Credit Modeling** theories, and the *risk* and *return* are taken to the account during the calculation of the *Cos of Capital* in *Ideas At Work (2006)*:

1. **Cost of Capital** is calculated using the *Weighted Average Cost of Capital (WACC)* model, which includes the following financial variables and ratios: *Levered Beta, Debt/Total Capitalization, Tax Rate, Unlevered Beta, Targeted Capital Structure, Risk Free Rate, Market Risk Premium, Spread over Risk Free Rate*. The *Weighted Average Cost of Capital (WACC)* is the weighted average of the marginal costs of all sources of capital. The formula for estimating *WACC* is as follows in *Schnoor (2006)*:

$$WACC = K_d(1-T)D/V + K_e E/V + K_p P/V$$

where:

$K_d$ = the *pre-Tax Cost of Debt*;

$T$ = the *Marginal Tax Rate* of the entity being valued;

$D/V$ = the Long-term target *Net Debt* to *Total Capitalization*;

$K_e$ = the market-determined *Cost of Equity Capital*;

$E/V$ = the Long-term target *Market Value of Equity* to *Total Capitalization*;

$K_p$ = the *Cost of Traditional Preferred Stock*;

$P/V$ = the Long-term target *Market Value of Preferred Stock* to *Total Capitalization*.

2. **Cost of Equity** is calculated using the *Capital Asset Pricing Model (CAPM)*, which includes the following financial variables and ratios: *Beta = Firm Specific Risk / Market Risk, Cost of Equity = Risk Free Rate + Beta, Multifactor Models of Asset Returns*. In *CAPM* theory in *Jarrow (1988), Lintner (1965), Sharpe (1964), Sharpe, Alexander, Bailey (1999)*, the **beta is a measure of risk**: *a measure of stock price volatility relative to the overall benchmark market index. The **beta** changes from 0 to 2 (beta=0, risk=0; beta=1, then risk=average market risk (a stock moves up or down in the same proportion as the overall market); beta=2, then risk=well above average market risk)*. The company's *Cost of Equity*, **Ke**, is calculated using the **Capital Asset Pricing Model (CAPM)** in *Schnoor (2006)*:

$$K_e = R_f + \beta^* \ \textbf{(Market Risk Premium)}$$

where:

$K_e$ = the market-determined *Cost of Equity Capital*;

$R_f$ = the *Risk Free Rate*;

*β* = the company's *beta*. The **beta** is a measure of stock price volatility relative to the overall benchmark market index. In other words, the **beta** is the price volatility of a financial instrument relative to the price volatility of a market or index as a whole. Beta is most commonly used with respect to equities. A high-beta instrument is riskier than a low-beta instrument. If a stock moves up or down in the same proportion as the overall market, it has a *Beta* of *1.0*. A stock with *Beta* of *1.2* is considered riskier than the overall market. *Higgins (2007)* states that the **beta** can also be considered as an angle of incline:

$$\beta = \frac{P_{jm} y_i}{y_m}$$

where $P_{jm}$ is the non-diversified risk.

*Bernanke (2009)* specifies the four main categories of risk to consider by the banks to satisfy the *Basel III capital requirements* in *Basel Committee on Banking Supervision (2006, 2009)*:

1. *Market Risk*;
2. *Credit Risk*;
3. *Operational Risk*;
4. *Rollover Risk*.

The additional categories of risk may include in *Ledenyov V O, Ledenyov D O (2012)*:

1. *Transaction risk*;
2. *Foreign exchange risk*;
3. *Reputation risk*;
4. *Emerging markets risk*;
5. *Environmental risk*;
6. *Geopolitical risk*.

Also, let us note that the risk management is extensively used in the corporate capital budgeting and allocation models in the process of critically important decision-making about the capital distribution, using the following methods in *Shinoda (2010)*:

1. *Net Present Value* (*NPV*) method;
2. *Internal Rate of Return* (*IRR*) method;
3. *Simple Payback Period* (*SPP*) method;
4. *Discounted Payback Period* (*DPP*) method;
5. *Accounting Rate of Return* (*ARR*) method, such as *Return on Investment (ROI)*, and
6. *Real Option* (*RO*) method.

Let us consider the appropriate modern approaches to model the volatility and evaluate the market risk. The **Autoregressive Conditional Heteroskedasticity (ARCH)** model in *Engle (1982, 2003)* is used in the field of statistical modeling of volatility in *Barone-Adesi,*

*Giannopoulos, Vosper (1999); McNeil A and Frey R (2000); Nelson D B*. The *ARCH* enables to model the financial and economic variables, such as the interest rates and equity prices, by performing the *Monte Carlo simulation,* using the stochastic differential equations (*SDE*). The **Generalized Autoregressive Conditional Heteroskedasticity (GARCH)** performs the modeling over the big window of sequential events, using the **weighted averages** and giving more weight to the recent events and less weight to the distant events in *Bollerslev (1986)*. The *GARCH* volatility is proportional to the **Value at Risk (VaR)**. *Engle (2003)* emphasized that the *GARCH* model presents the **theory of dynamic volatilities**. We apply the integrative thinking in *Martin (2005-2009)* to analyze the *GARCH* imperfections and limitations and conclude that the new volatility model, based on the **theory of nonlinear dynamic volatilities,** must be developed to describe the volatilities in the real world nonlinear dynamical financial system, going from the econophysical analysis of **appearing nonlinearities during the interaction between the various business cycles** in *Ledenyov D O, Ledenyov V O (2012)*. We state that the simple nonlinear financial systems do not possess the simple dynamical properties, because the nonlinear dynamical systems are usually characterized by the self-interactions, self-organizations, spontaneous emergence of order, dissipative structures and nonlinear cooperative phenomena in *Uechi, Akutsu (2012); Mosekilde (1996, 1996-1997); Kuznetsov (2001, 1996-1997); Kuznetsov A P, Kuznetsov S P, Ryskin N M (2002)*. For example, in the case of nonlinear financial system, the series of sequential events can have a point of crisis that could magnify the small changes and lead to the big nonlinearities, caused by the *Butterfly effect* with the strong dependence on the initial conditions of the nonlinear financial system. In other words, the simple nonlinear financial system has the very complex nonlinear dynamical properties, which can only be described, using the **Dynamic Chaos** theory. It means that the *GARCH* model together with the *high frequency volatility model* and *high dimension volatility model* in *Engle (1982, 2003)* do not take to the account the various nonlinearities during the statistical modeling of financial volatility**,** hence they are not accurate and have to be revised or disregarded.

We believe that it is necessary to improve the financial responsibility and risk management in the time of high volatilities in the capital markets in *Cameron (2008)*. Therefore, we propose the new *theory of nonlinear dynamic volatilities* for the accurate statistical characterization and modeling of financial volatility, using the *nonlinear stochastic differential equations (NSDE)*. Our new **nonlinear dynamic chaos (NDC) volatility model**, which is based on both the **nonlinear dynamic volatilities theory** and the **dynamic chaos theory**, applies the econophysical analysis to predict the *nonlinearities appearance during the interaction between the various business and financial cycles with the different amplitudes, periods, phases*. The **NDC volatility model** proved to be a useful model, which can be used to determine the *VaR* precisely and perform the investment portfolio management accurately. The comprehensive discussion on the *NDC volatility model* is beyond the scope of this short article and it can be found in our other research publications.

We conclude with the remark that we proposed the new *theory of nonlinear dynamic volatilities* and the new *nonlinear dynamic chaos (NDC) volatility model* for the statistical modeling of financial volatility with the aim to set the *Value at Risk (VaR)* and manage the financial portfolio, applying the following techniques: 1) *Risk aggregation*, which aims to get rid of non-systematic risks with diversification and 2) *Risk decomposition*, which tackles risks one by one.

Authors are very grateful to the *Yukawa Institute for Theoretical Physics* at *Kyoto University, Kyoto, Japan* and especially to *Profs. Hideaki Aoyama, Kyoto University; Yoshi Fujiwara, University of Hyogo; Hiroshi Iyetomi, University of Tokyo; Aki-Hiro Sato, Kyoto University* for a kind opportunity to get an open access and analyze the research papers, presented at the *YITP* workshop on "*Econophysics 2011 — The Hitchhiker's Guide to the Economy*." We appreciate the *Graduate School of Economics and Business Administration at Hokkaido University, Sapporo, Hokkaido, Japan* for giving us a wonderful opportunity to conduct the research on the highly innovative research papers, written by the *Japanese* scientists. We thank *Prof. Shigetoshi Ohshima* from *Graduate School of Science and Engineering* at *Yamagata University, Yonezawa, Japan* for his strong research interest in the origin of nonlinearities in *High Temperature Superconductors (HTS)* in *Ultra High Frequency (UHF) Electromagnetic Fields*.

The first author appreciates *Prof. Janina E. Mazierska, Electrical and Computer Engineering Department, School of Engineering and Physical Sciences, James Cook University, Australia* for an opportunity to make the advanced innovative research on the nonlinear dynamic microwave resonant systems in the field of superconducting electronics during more than 12 years.

The second author appreciates *Profs. Roger L. Martin* and *John C. Hull* for the presented opportunity to learn more about the integrative thinking and risk management in the finances and economics in North America in *1998-1999* and in *2005-2006*. *Profs. Roger L. Martin* and *John C. Hull* valuable advices on the *Bloomberg terminal* operation to obtain the *Levered Beta* as one of the important inputs during the calculation of the *Weighted Average Cost of Capital* in the various *North American* companies at the *Trading Lab at the Rotman School of Management* at *University of Toronto, Toronto, Canada* in *2005-2006* are also acknowledged.

*Lionel Barber, Editor-in-Chief, Financial Times* is appreciated for the regular exchange by interesting opinions on the financial topics as well as his kind encouragements, including the numerous invitations to discuss the complex issues in the fields of finances and economics with more than one hundred global leaders, economists, financiers, professors and journalists in the *FT* in London in the *UK* in recent years.

[*]) This condensed version of our research article is submitted to *The Financial Times, The Bodley Head and The Random House first annual essay competition* in London in the *UK* in 2012.

*E-mail: dimitri.ledenyov@my.jcu.edu.au